\documentclass{aa}
\usepackage{graphicx}
\usepackage{txfonts}

\newcommand{\be}{\begin{equation}}
\newcommand{\ee}{\end{equation}}
\newcommand{\bea}{\begin{eqnarray}}
\newcommand{\eea}{\end{eqnarray}}

\def\mnras{MNRAS}
\def\apj{ApJ}

\def\aea{A\&A}

\def\aas{A\&AS}

\def\la{\mathrel{\hbox{\rlap{\hbox{\lower4pt\hbox{$\sim$}}}\hbox{$<$}}}}
\def\ga{\mathrel{\hbox{\rlap{\hbox{\lower4pt\hbox{$\sim$}}}\hbox{$>$}}}}

\def\la{\mathrel{\hbox{\rlap{\hbox{\lower4pt\hbox{$\sim$}}}\hbox{$<$}}}}
\def\ga{\mathrel{\hbox{\rlap{\hbox{\lower4pt\hbox{$\sim$}}}\hbox{$>$}}}}

\def\aea{A\&A}
\def\apj{ApJ}

\begin{document}

\title{The X-ray spectrum of \object{$\delta$ Orionis} observed
by {\it LETGS} aboard {\it Chandra}
}

\author{A.J.J.\,Raassen\inst{1,2}
\and A.M.T. Pollock\inst{3}
}

\institute{
SRON Netherlands Institute for Space Research, Sorbonnelaan 2, 3584 CA Utrecht, The Netherlands 
\and 
Astronomical Institute "Anton Pannekoek", Science Park 904, 1098 XH Amsterdam, University of Amsterdam, The Netherlands
\and
European Space Agency XMM-Newton Science Operations Centre, European Space Astronomy Centre, Apartado 78, Villanueva de la Ca\~nada, 28691 Madrid, Spain
   }

\offprints {A.J.J.Raassen,\\
\email a.j.j.raassen@sron.nl}
\authorrunning{A.J.J.Raassen \& A.M.T. Pollock}
\titlerunning{The {\it LETGS} X-ray spectrum of \object{$\delta$ Orionis}}

\date{26~November~2012}


\abstract{}{We analyze the high-resolution X-ray spectrum of the supergiant O-star
  \object{$\delta$ Orionis} (O9.5II) with line ratios of He-like ions and a thermal plasma model,
 and we examine its variability.}
 {The O-supergiant {\object$\delta$ Ori} was observed in the wavelength range 5--175~\AA\ by
  the X-ray detector HRC-S in combination with the grating LETG aboard {\sl Chandra}.
  We studied the He-like ions in combination with the UV-radiation field 
to determine local plasma temperatures and to establish the
distance of the X-ray emitting ions to the stellar surface. 
 We measured individual lines by means of Gaussian profiles, folded through the response matrix,
 to obtain wavelengths, line fluxes,
 half widths at half maximum (HWHM) and line shifts to characterize the plasma.
We consider
 multitemperature models in collisional ionization equilibrium (CIE) 
to determine temperatures, emission measures, and abundances.
  }
{
Analysis of the He-like triplets extended to
  \ion{N}{\sc vi} and \ion{C}{\sc v} implies ionization
  stratification with the hottest plasma to be found
within a few stellar radii 3~$R_*$ (\ion{Mg}{\sc xi}) and the coolest farther out,
far beyond the acceleration zone, up to 49~$R_*$ (\ion{N}{\sc vi}) and 75~$R_*$ (\ion{C}{\sc v}). 
The observed temperatures cover a range from about 0.1 to 0.7~keV, i.e., 1--8~MK.
  The X-ray luminosity ($L_x$) is $\sim 1.5\times 10^{32}$erg/s in the range from 0.07 to 3~keV
  covered by {\it LETGS}. 
Velocity widths of about 1040~km~s$^{-1}$ 
have been determined. 
}{}

\maketitle

\section{Introduction}
Since the launch of the first X-ray observatories, it has been known that O-type stars 
and early B-type stars are X-ray sources. Harnden et al. (1979),
Seward et al. (1979), and
Cassinelli et al. (1981) described the first observations with {\it Einstein}. The investigations were  continued with
{\it ASCA} (Corcoran et al. 1994) and {\it ROSAT} (Haberl \& White 1993, Bergh{\" o}fer et al. 1996).
Although the origin 
of the emission was initially a matter of debate, it is now generally believed that instabilities and clumping in the 
strong line-driven stellar wind is the mechanism that heats small parts of the wind 
(e.g., Lucy \& Solomon 1970; Lucy \& White 1980; Lucy 1982; Owocki et al. 1988; Dessart \& Owocki 2005). The
launches of {\it Chandra} and {\it XMM-Newton} in 1999 with high-resolution instruments aboard have offered the
possibility to study the X-ray spectra in more detail, resulting in, e.g., 
 estimates of line ratios, line profiles, temperatures, emission measures, and abundances
(Waldron \& Cassinelli 2001, 2007; Cohen et al. 2006; Kahn et al. 2001; Raassen et al. 2008).
Generally, the overall spectra have been described by means of collisional ionization equilibrium (CIE) models with several temperature
components, often complemented by analysis of individual line profiles, line fluxes, and line flux ratios of He-like ions.
Based on {\it RGS} observations of \object{$\zeta$ Ori}, Pollock
(2007) suggests an alternative mechanism due to the non-equilibrium ion-ion interactions
far out in the wind.
Since extending the high-resolution coverage to the longer wavelengths available with the {\it Chandra LETGS}
should provide new constraints on theory,
it has been proposed to record the spectrum of \object{$\delta$ Ori}, the softest of the X-ray brightest O-stars observed with {\it ROSAT}.
This paper discusses the {\it LETGS} data obtained in the context of individual line diagnostics 
to obtain plasma characteristics, 
as well as overall thermal models.

The O-star \object{$\delta$ Ori} (HD~36486) is the western star in the belt of Orion, named Mintaka, with a parallax of 4.71(0.58) mas, 
i.e., at a distance of 212~pc (SIMBAD: Van Leeuwen 2007)
\footnote{\textrm {http://simbad.u-strasbg.fr/simbad/sim-fid}}. The position in J2000 coordinates is 
\hbox{RA=05 32 00.400} and \hbox{Dec=$-$00 17 56.74}.
It is a bright X-ray source and
suffers lower interstellar absorption than \object{$\zeta$ Ori}. It is part of a
triple star system, containing of \object{$\delta$ Ori}A, \object{$\delta$ Ori}B, and \object{$\delta$ Ori}C.
Here we focus on the ``single star'' \object{$\delta$ Ori}A, which in itself is also a triple star system (Aa1, Aa2,
and Ab). The tertiary star (Ab) is too far away to play a role in our observation. The primary (Aa1), a supergiant,
is of spectral type O9.5II, with a temperature of 33000K. The secondary (Aa2) is of spectral
type B0.5III, with a temperature of 27000K (Voels et al.1989 and Tarasov et al. 1995). They form a binary system with a period of 5.7325 days (Harvey et al. 1987). 
Different stellar mass values are given in the literature. Harvin et al. (2002) give
 $M_p = 10.3M_{\odot}$ as average 
with $R_p = 11R_{\odot}$  
and for the secondary star $M_s = 5.2M_{\odot}$ and $R_s = 4R_{\odot}$, while Mayer et al. (2010) suggest 
values of $M_p = 25M_{\odot}$ with $R_s = 16-17R_{\odot}$. The primary and secondary stars are separated by 33$R_{\odot}$, equivalent 
to 3$R_p$. The properties of \object{$\delta$ Ori}, which we have applied, are given in Table~\ref{table-quantities}.

\begin{table}[h!]
\caption{Adopted properties of \object{$\delta$ Ori}. }
\begin{center}
\begin{tabular}{l@{\ }c@{\ }}
\hline
\hline
Property               &~~~~~~~~ \object{$\delta$ Ori}\\
\hline
Distance from Earth(pc)&~~~~~~~~212$^a$ \\
Spectral type           &~~~~~~~~O9.5II$^b$\\
Radius($R_{\odot}$)       &~~~~~~~~11$^b$\\
Mass($M_{\odot}$)         &~~~~~~~~10.3$^b$\\
$T_{eff}$(K)              &~~~~~~~~33000$^c$\\
Terminal wind velocity(km~s$^{-1})$&~~~~~~~~2000$^d$\\
$\dot{M}(M_{\odot}$yr$^{-1}$)&~~~~~~~~1.07$\times$10$^{-6}$$^d$\\
\hline
\end{tabular}
\label{table-quantities}
\end{center}
\begin{flushleft}
{
Notes:\\
$a$: Van Leeuwen (2007)\\
$b$: Miller et al.(2002), taken from Harvin et al. (2002).\\
$c$: Miller et al.(2002), taken from Voels et al. (1989).\\
$d$: Miller et al.(2002), taken from Lamers \& Leitherer (1993).\\

}
\end{flushleft}
 
\end{table}

The spectrum of \object{$\delta$ Ori} was studied in
X-rays with earlier satellites by Cassinelli \& Swank (1983), Haberl \& White (1993), and Corcoran et al. (1994)
and most
recently at high resolution by Waldron \& Cassinelli (2007) and Miller et al. (2002) with the {\it Chandra HETGS}.
They focused on line profiles and line widths, together with ratios of forbidden and intercombination lines in
He-like ions.

The set up of this paper is as follows.
The emphasis of the paper is on determining the location of the X-ray emitting plasma 
and its temperature in relation
to the stellar surface and the UV-radiation field. Therefore we need individual line fluxes.
In Sect.~2 the observation and the lightcurve are discussed.
We show individual line fluxes, line wavelengths, and line broadenings in Sect.~3. Based on the 
individual line fluxes of He-like ions, the location relative to the stellar surface,  
as well as the temperature of the plasma, is derived. They are given in Sect.~4.
We discuss a multitemperature model fit to the total spectrum with a CIE-model 
in Sect.~5.
 As suggested by the light curve, we also make a time split, 
dividing the total
observation of 98~ks into the first part of 49~ks and the second part of 49~ks.

\section{Observation and lightcurve}
\subsection{Data and pipelining}
The X-ray spectrum of \object{$\delta$ Ori} was obtained by means of {\it HRC-S} 
in combination with the Low-Energy-Transmission-Grating ({\it LETG}) 
on board {\it Chandra} on November 9, 2007 during 97~ks (JD=54413.92705-54415.05066).
The log of the observation is shown in Table~\ref{log-observation}. 

The data were pipelined with CIAO3.4 with CALDB
version~3.4.1 and prepared for SPEX by the programs {\it extract} and {\it crele} developed at SRON.
The spectrum derived from this observation is the sum of the +1 and -1 order. The background,
which is quite dominant in the higher wavelength area of {\it LETG} spectra, has been subtracted. The spectrum contains 31500 counts in the
spectral range from 5 to 175~\AA.

\begin{figure}[h!]
\hbox
{\includegraphics[width=\columnwidth]{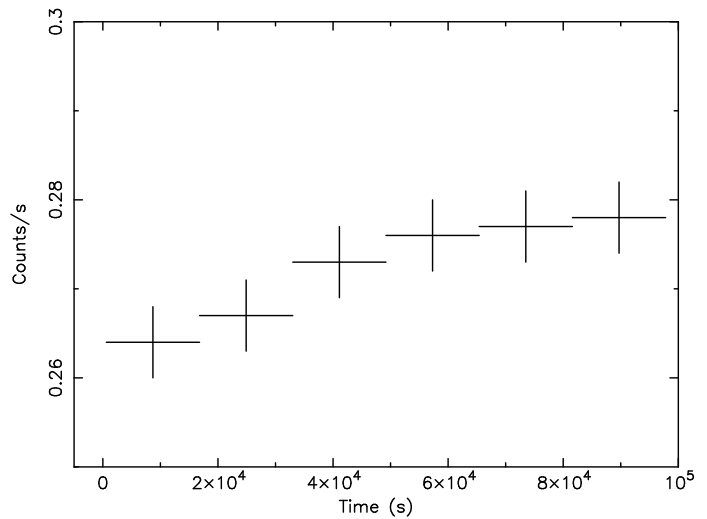}}
\caption{The zeroth-order lightcurve of \object{$\delta$ Ori}, observed with LETGS, 
and rebinned to time bins of 15700~seconds. The curve indicates a minimum at the beginning of the observation
and a flattening of the increase at the end of the observation.}
\label{lightcurve}
\end{figure} 

\begin{table*}[ht!]
\caption{Observation log of the data of \object{$\delta$ Ori}.}
\begin{center}
\begin{tabular}{l@{\ }l@{\ }l@{\ }l@{\ }l@{\ }c@{\ }c@{\ }c@{\ }}
\hline
\hline
Obs ID ~~~&Instrument        ~~~&  Grating~~~&Start Date~~~ &End Date~~~ &Duration(ks)~~~&Wavelength range$^a$\\
\hline
7416 ~~~&HRC-S             ~~~&  LETG   ~~~&2007-11-09T10:14:57~~~&2007-11-10T13:12:57~~~&97.08~& 5-175~\AA\ \\
  &                   &            &JD~54413.92705     ~~~&JD~54415.05066     ~~~&      &            \\
\hline
\end{tabular}
\label{log-observation}
\end{center}
\begin{flushleft}
{
Notes:\\
$a$: Range also used during the fitting procedure.\\
}
\end{flushleft}
 
\end{table*}

\subsection{Zeroth-order lightcurve}
In other wavelength ranges (optical and IUE) fluctuations due to eclipsing have been observed
in the lightcurve of \object{$\delta$ Ori}. These changes are about .12m for the primary
eclipse and .07m for the 
secondary eclipse (Harvin et al. 2002). This corresponds to a decrease of about 11\% during the
primary eclipse
and of about 6\% during the secondary eclipse.
Based on the zeroth-order image a lightcurve has been constructed.
Figure~\ref{lightcurve} shows the zeroth-order lightcurve of this observation of \object{$\delta$ Ori}. 
The total number of counts 
in this zeroth-order image is 26350 counts. No periodicity could be established in the lightcurve, 
as this observation is too short to cover several stellar revolutions. 
However, after rebinning to time bins of about 16~ks the lightcurve shows some variation in 
count rate from 0.264(0.004) to 0.278(0.004) counts/s during this observation. This is a change of about 5\%. 
During the observation the phase of the binary system was 0.83-0.02, derived from Harvin et al. (2002), or 0.87-0.07, as
derived from Mayer et al. (2010). 
This implies that the secondary is in front of the primary during the observation.
In later sections we investigate a possible relation between the behavior of the lightcurve and the 
observed spectrum by cutting the observation into a first part from 0 to 49~ks
and a second part from 49 to 98~ks.

\section{Individual line fluxes}
\begin{figure}[h!]
\hbox
{\includegraphics[width=\columnwidth]{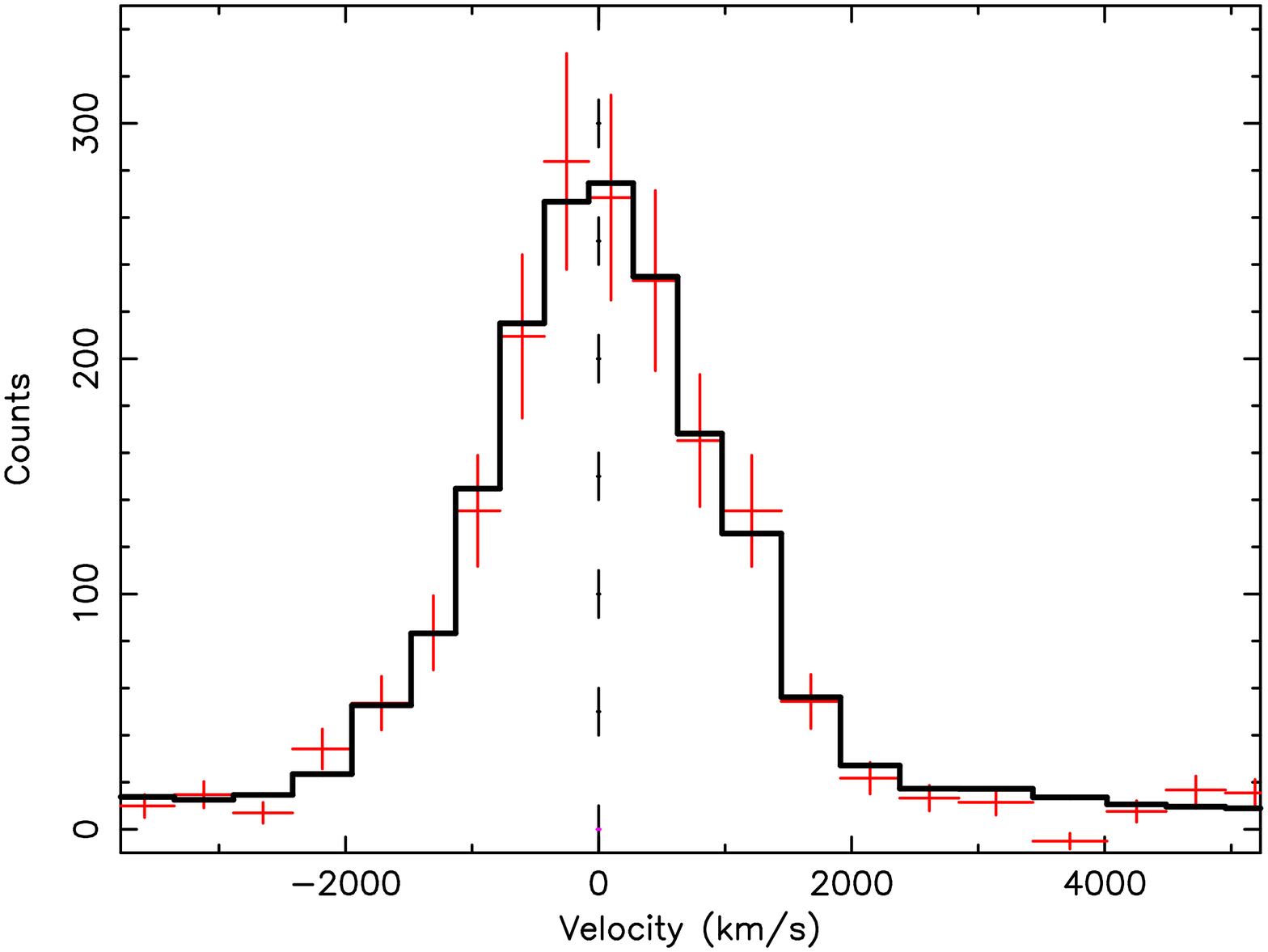}}
{\includegraphics[width=\columnwidth]{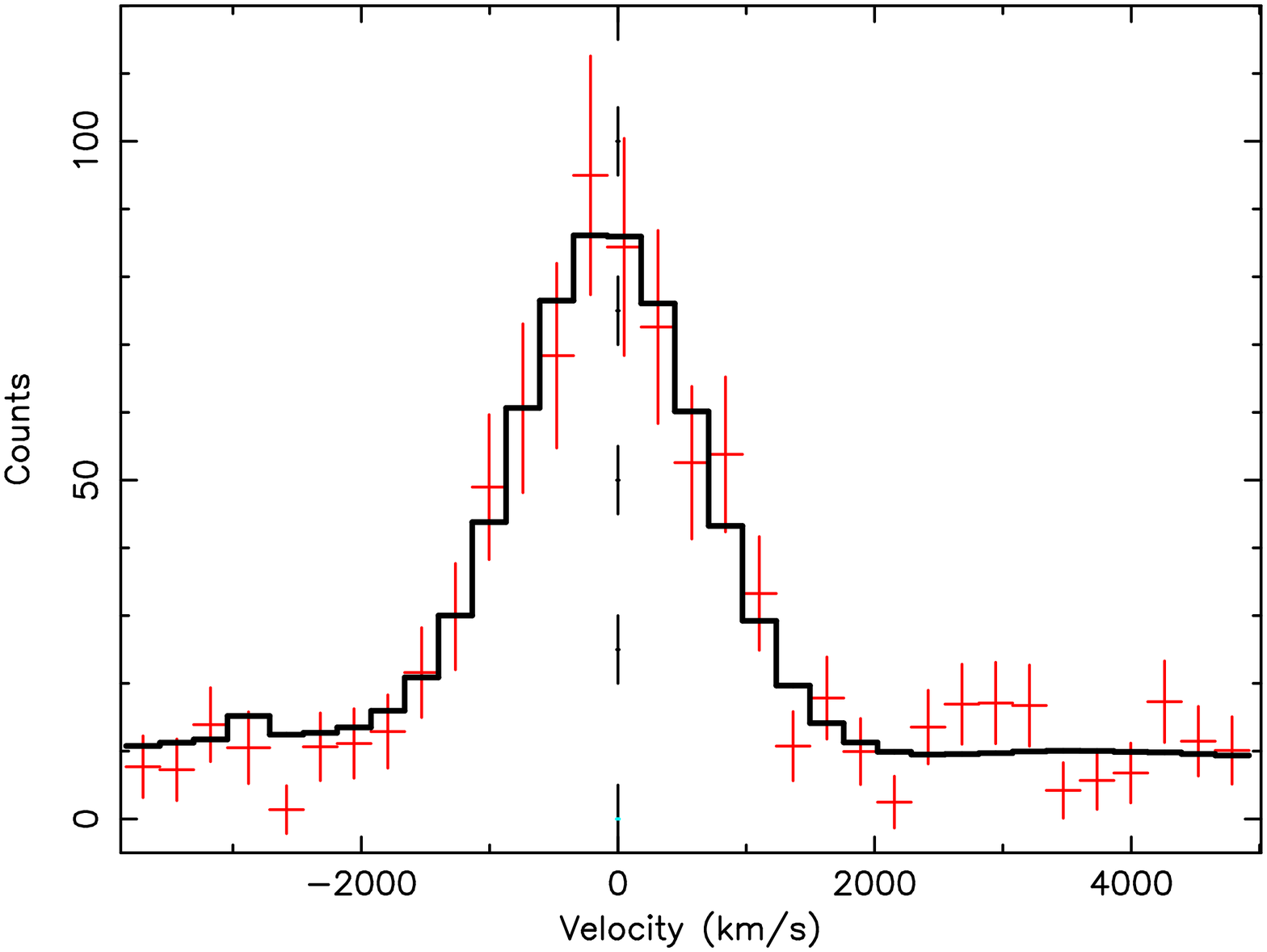}}
\caption{Line shapes of the \ion{O}{\sc viii} line at 18.97~\AA\ and the \ion{C}{\sc vi} line at 33.74~\AA\ 
in the spectrum of \object{$\delta$ Ori}. Data are in red/gray, the fitted Gaussian model line folded through the
response matrix is given in black. The rest wavelength is also shown. Keep in mind that the width is also 
influenced by the instrumental line spread function.}
\label{lineprofile} 
\end{figure}

The individual line fluxes of the {\it LETGS} spectrum of \object{$\delta$~Ori} were measured. 
For each observed line a Gaussian profile was folded through the response matrix and fit to the observed line profile, establishing the line positions, 
line fluxes, as well as the linewidths. A powerlaw was added to describe the continuum near the 
line features. The line profiles are symmetric. This is shown in Fig.~\ref{lineprofile} for the O~VIII and
C~VI lines. This agrees with observations of $\delta$ Ori by Miller et al. (2002), when applying {\it HETGS}. 
Seeming asymmetries in the data are 
due to asymmetries in the instrumental line spread function (LSF). It also influences the observed widths of the lines. 
The situation in $\zeta$ Ori is different. In that star the lines are asymmetric and broadened by themselves as
shown by Pollock (2007). 

The measured wavelengths (\AA) and line fluxes (in units 10$^{-4}$~photons~cm$^{-2}$s$^{-1}$
and in  units 10$^{-13}$~ergs~cm$^{-2}$s$^{-1}$) are collected in Table~\ref{lineflux}, together with the theoretical
wavelengths from laboratory measurements collected by Kelly (1987) and the {\it NIST4}~database 
\footnote{\textrm {http://www.nist.gov/pml/data/asd.cfm}}. The stronger lines have a low value for the (1~$\sigma$) uncertainty. 
Their uncertainties are a few m\AA. 

\begin{table}[ht!]
  \caption[]{Line positions and line fluxes in $\delta$ Ori. 
 }
  \label{lineflux}
\begin{tabular}{l@{\ }r@{\ }c@{\ }c@{\ }c@{\ }}		 
		   \hline \hline

ion          & $\lambda_0^a$& $\lambda_{\rm obs}^b$&flux$^c$&flux$^d$\cr 
             & [\AA]        &[\AA]                 &        &   \cr 
\hline
S\,{\sc xv}    & 5.038+.066&4.977(.103)$^f$&~~~ 0.27(.15)&~~~~1.07\cr
Si\,{\sc xiv}  & 6.182 &  6.168(.029) &~~~ 0.05(.03)&~~~~0.14\cr
Si\,{\sc xiii}  & 6.648 &  6.635(.007) &~~~ 0.14(.03)&~~~~0.43\cr
Si\,{\sc xiii}& 6.688+.704 &6.698(.011)&~~~ 0.08(.03)&~~~~0.24\cr
Si\,{\sc xiii}  & 6.740 &  6.751(---)  &~~~ 0.01(.01)&~~~~0.03\cr
Mg\,{\sc xii}   & 8.422 &  8.423(.009) &~~~ 0.09(.03)&~~~~0.21\cr
Mg\,{\sc xi}    & 9.168 &  9.165(.008) &~~~ 0.18(.04)&~~~~0.39\cr
Mg\,{\sc xi}    & 9.231 &  9.229(.010) &~~~ 0.12(.04)&~~~~0.25\cr
Mg\,{\sc xi}    & 9.314 &  9.315(.011) &~~~ 0.08(.03)&~~~~0.17\cr
Ne\,{\sc x}     & 9.710 &  9.716(.019) &~~~ 0.08(.04)&~~~~0.17\cr
Ne\,{\sc x}     &10.240 & 10.238(.012) &~~~ 0.08(.03)&~~~~0.15\cr
Fe\,{\sc xix}   &10.640 & 10.647(.008) &~~~ 0.11(.04)&~~~~0.21\cr
Fe\,{\sc xvii}  &11.130 & 11.136(.019) &~~~ 0.05(.03)&~~~~0.09\cr
Fe\,{\sc xvii}  &11.251 & 11.274(.016) &~~~ 0.09(.04)&~~~~0.16\cr
Fe\,{\sc xviii} &11.530 & 11.552(.009) &~~~ 0.16(.04)&~~~~0.28\cr
Ne\,{\sc x}     &12.134 & 12.131(.003) &~~~ 1.00(.08)&~~~~1.66\cr
Fe\,{\sc xxi}   &12.286 & 12.271(.008) &~~~ 0.28(.05)&~~~~0.45\cr
Ni\,{\sc xix}   &12.430 & 12.437(.015) &~~~ 0.08(.04)&~~~~0.13\cr
Ni\,{\sc xix}   &12.654 & 12.662(.022) &~~~ 0.07(.04)&~~~~0.11\cr
Fe\,{\sc xx}    &12.818+.847 & 12.845(.009) &~~~ 0.25(.05)&~~~~0.39\cr
Fe\,{\sc xx}    &12.946+.978 & 12.973(.014) &~~~ 0.12(.04)&~~~~0.18\cr 
???             & ---   & 13.257(.016) &~~~ 0.10(.04)&~~~~0.16\cr
Ne\,{\sc ix}    &13.447 & 13.450(.006) &~~~ 1.45(.15)&~~~~2.15\cr
Ne\,{\sc ix}    &13.553 & 13.542(.007) &~~~ 1.10(.14)&~~~~1.61\cr
Ne\,{\sc ix}    &13.700 & 13.723(.011) &~~~ 0.27(.05)&~~~~0.39\cr
Fe\,{\sc xvii}  &13.823 & 13.824(.009) &~~~ 0.42(.06)&~~~~0.60\cr
Ni\,{\sc xix}   &14.040+.081 & 14.052(.008) &~~~ 0.31(.05)&~~~~0.43\cr
Fe\,{\sc xviii} &14.202+.255 & 14.209(.005) &~~~ 0.76(.11)&~~~~1.06\cr
Fe\,{\sc xviii} &14.344+.361 & 14.364(.008) &~~~ 0.34(.05)&~~~~0.48\cr
Fe\,{\sc xvii}  &15.015 & 15.011(.002) &~~~ 2.54(.11)&~~~~3.36\cr
O\,{\sc viii}   &15.176 & 15.175(.012) &~~~ 0.50(.10)&~~~~0.64\cr
Fe\,{\sc xvii}  &15.260 & 15.264(.008) &~~~ 1.03(.11)&~~~~1.34\cr
???             & ---   & 15.844(.013) &~~~ 0.20(.12)&~~~~0.24\cr
O\,{\sc viii}   &16.007 & 16.002(.006) &~~~ 0.89(.11)&~~~~1.11\cr
Fe\,{\sc xvii}  &16.073 & 16.092(.014) &~~~ 0.34(.09)&~~~~0.42\cr
Fe\,{\sc xvii}  &16.229 & 16.220(---)  &~~~ $\la$0.07&~~~~0.08\cr
Fe\,{\sc xix}   &16.284 & 16.296(.020) &~~~ 0.06(.03)&~~~~0.07\cr
Fe\,{\sc xvii}  &16.777 & 16.775(.003) &~~~ 1.39(.09)&~~~~1.62\cr
Fe\,{\sc xvii}  &17.054 & 17.054$^e$   &~~~ 2.48(.18)&~~~~2.89\cr
Fe\,{\sc xvii}  &17.097 & 17.097$^e$   &~~~ 1.02(.16)&~~~~1.18\cr
O\,{\sc vii}    &17.768 & 17.763(.012) &~~~ 0.16(.05)&~~~~0.18\cr
O\,{\sc vii}    &18.627 & 18.628(.008) &~~~ 0.58(.07)&~~~~0.62\cr
O\,{\sc viii}   &18.969 & 18.968(.002) &~~~ 6.60(.18)&~~~~6.91\cr
N\,{\sc vii}    &20.910 & 20.894(---)  &~~~ 0.06(---)&~~~~0.05\cr
O\,{\sc vii}    &21.602 & 21.614(.003) &~~~ 6.29(.24)&~~~~5.81\cr
O\,{\sc vii}    &21.804 & 21.793(.003) &~~~ 5.57(.23)&~~~~5.07\cr
O\,{\sc vii}    &22.101 & 22.101$^e$   &~~~ 0.25(.09)&~~~~0.22\cr
Ca\,{\sc xiv}+S\,{\sc xiv}&24.133+.285& 24.167(.056) &~~~ 0.26(.11)&~~~~0.21\cr
N\,{\sc vii}    &24.781 & 24.788(.011) &~~~ 1.38(.15)&~~~~1.11\cr
Ar\,{\sc vii}   &25.684 & 25.705(.030) &~~~ 0.46(.12)&~~~~0.36\cr
C\,{\sc vi}     &26.357 & 26.356(---)  &~~~ 0.26(---)&~~~~0.20\cr
C\,{\sc vi}     &26.990 & 26.990(.047) &~~~ 0.30(.13)&~~~~0.23\cr
C\,{\sc vi}     &28.466 & 28.462(.017) &~~~ 0.57(.15)&~~~~0.40\cr
N\,{\sc vi}     &28.787 & 28.763(.028) &~~~ 1.27(.20)&~~~~0.88\cr
N\,{\sc vi}     &29.084 & 29.065(.037) &~~~ 0.86(.18)&~~~~0.58\cr
N\,{\sc vi}     &29.534 & 29.475(.060) &~~~ 0.36(.14)&~~~~0.24\cr

\hline
\end{tabular}
\end{table}
\begin{table}[h!]
\begin{tabular}{l@{\ }r@{\ }c@{\ }c@{\ }c@{\ }}		 
		   \hline \hline

ion          & $\lambda_0^a$& $\lambda_{\rm obs}^b$&flux$^c$&flux$^d$\cr 
             & [\AA]        &[\AA]                 &        &   \cr 
\hline
C\,{\sc vi}     &33.736 & 33.728(.008) &~~~ 4.88(.39)&~~~~2.88\cr
Ca\,{\sc xi}    &35.212 & 35.175(.078) &~~~ 0.53(.18)&~~~~0.30\cr
Ca\,{\sc xi}    &35.576 & 35.580(.054) &~~~ 0.62(.18)&~~~~0.35\cr
S\,{\sc xii}&36.563+.398& 36.500(.036) &~~~ 0.74(.20)&~~~~0.40\cr
Si\,{\sc xiii}+S\,{\sc xi} &37.786+.773& 37.787(.040) &~~~ 0.83(.23)&~~~~0.43\cr
S\,{\sc xi} &39.240+.323& 39.234(.036) &~~~ 1.43(.29)&~~~~0.72\cr
C\,{\sc v}      &40.268 & 40.240(.020) &~~~ 2.74(.74)&~~~~1.35\cr
C\,{\sc v}      &40.729 & 40.730$^e$   &~~~ 1.96(.53)&~~~~0.97\cr
C\,{\sc v}      &41.472 & 41.474(.024) &~~~ 0.66(.32)&~~~~0.32\cr
Mg\,{\sc x}+Si\,{\sc xii}&44.050+.020& 44.031(.022) &~~~ 0.85(.13)&~~~~0.39\cr
Si\,{\sc ix}+{\sc xii}  &44.249+.165& 44.265(.017) &~~~ 0.92(.13)&~~~~0.41\cr
Si\,{\sc xii}  &45.520+.680&45.533(.033) &~~~ 0.20(.09)&~~~~0.09\cr
Si\,{\sc xi}   &46.300+.410&46.308(.062) &~~~ 0.42(.11)&~~~~0.18\cr
Mg\,{\sc x}+S\,{\sc ix} &47.280+.500&47.410(.050) &~~~ 0.58(.12)&~~~~0.24\cr
Ar\,{\sc ix}+Si\,{\sc xi}&49.180+.220&49.165(.025) &~~~ 1.10(.16)&~~~~0.44\cr
Si\,{\sc x }   &50.254+.305&50.284(.015) &~~~ 1.07(.21)&~~~~0.42\cr
Si\,{\sc x }   &50.524+.691&50.545(.031) &~~~ 1.21(.20)&~~~~0.48\cr
Si\,{\sc x }   &52.1784+.248&52.232(.038) &~~~ 0.72(.22)&~~~~0.28\cr
Fe\,{\sc xv}+Si\,{\sc ix}&52.911+.834&52.838(.044) &~~~ 1.04(.24)&~~~~0.39\cr
Fe\,{\sc xvi}  &54.710 & 54.700(.045) &~~~ 0.72(.19)&~~~~0.26\cr
Si\,{\sc ix }  &55.356+.272&55.340(.028) &~~~ 1.50(.22)&~~~~0.54\cr
Fe\,{\sc xv}   &59.404 & 59.350(.103) &~~~ 0.38(.15)&~~~~0.13\cr
???             & ---   &60.003(.078) &~~~ 0.40(.15)&~~~~0.13\cr
Si\,{\sc ix}+{\sc xiii} &61.050-.84& 61.416(.087)$^f$&~~~ 1.90(.14)&~~~~0.61\cr
Si\,{\sc ix}+{\sc xii} &68.148+.548& 68.328(.081) &~~~ 0.70(.22)&~~~~0.20\cr
Si\,{\sc viii}  &69.825 & 69.812(.055) &~~~ 0.81(.21)&~~~~0.23\cr

\hline
\end{tabular}
\begin{flushleft}
{
Notes:\\
$a$: $\lambda_0$ are the theoretical wavelength values from Kelly (1987).\\
$b$: $\lambda_{\rm obs}$ is the observed wavelength with the statistical 1$\sigma$ error
in parentheses.\\
$c$: in units 10$^{-4}$~photons~cm$^{-2}$s$^{-1}$\\
$d$: in units 10$^{-13}$~ergs~cm$^{-2}$s$^{-1}$\\
$e$: wavelength fixed at theoretical values\\
$f$: broad line\\
}
\end{flushleft}
\end{table}

\begin{table}[ht!]
  \caption[]{Higher-order lines and their line fluxes in the spectrum of $\delta$ Ori. 
 }
\label{higherOrderLines}
\begin{tabular}{l@{\ }r@{\ }r@{\ }r@{\ }r@{\ }}		 
		   \hline \hline

ion          & $\lambda_0^a$& $\lambda_{\rm obs}^b$& $\lambda_{\rm obs}$/order&flux$^c$\cr 
             & [\AA]        &[\AA]                 &[\AA]                 &          \cr 
\hline
Fe\,{\sc xvii}&15.015&30.020(.012)&~~15.010(.006)&~~~ 0.27(.09)\cr
Ne\,{\sc ix}&13.447+.553&40.545(~----~) &~~13.515(~----~)&~~~ 1.12(~----)\cr
Fe\,{\sc xvii}&15.015&45.004(.026)&~~15.001(.009)&~~~ 0.47(.10)\cr
Fe\,{\sc xvii}&17.054+.097&51.204(.067)&~~17.068(.022)&~~~ 0.50(.17)\cr
O\,{\sc viii}&18.969&56.936(.030)&~~18.978(.010)&~~~ 1.79(.24)\cr
O\,{\sc vii}&21.602+.804&65.171(.165)&~~21.724(.055)&~~~ 1.86(.52)\cr
C\,{\sc vi}&33.736&101.222(.348)&~~33.741(.116)&~~ 2.02(.88)\cr

\hline
\end{tabular}
\begin{flushleft}
{
Notes:\\
$a$: $\lambda_0$ are the theoretical wavelength values from Kelly (1987).\\
$b$: $\lambda_{\rm obs}$ higher-order observed wavelength with the statistical 1$\sigma$ error
in parentheses.\\
$c$: in units 10$^{-4}$~photons~cm$^{-2}$s$^{-1}$.\\
}
\end{flushleft}
\end{table}

\begin{figure*}[hb!]
\hbox
{\includegraphics[width=2.0\columnwidth,height=9.0truecm]{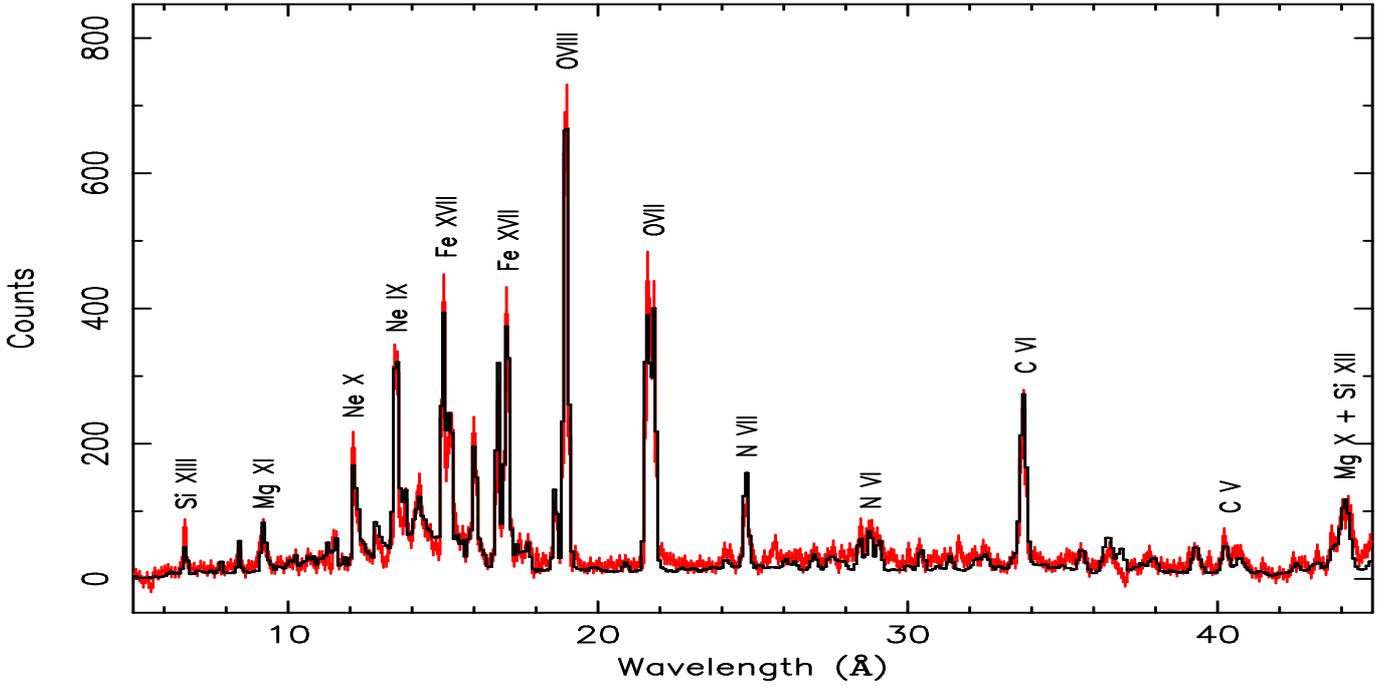}}
\caption{Part of the spectrum of \object{$\delta$ Ori}. The most dominant lines are indicated. Data
are in red (gray), the CIE-model in black)}
\label{spectrallines} 
\end{figure*} 

There are very few line features above
50~\AA, and they are often due to  higher-order lines, especially third-order lines,
 which have their strong first-order 
component between 10 and 35~\AA. The second-order lines of \ion{Fe}{\sc xvii} at 
16.777, 17.054, and 17.097~\AA\ contribute a 
 bit to the wings of the \ion{C}{\sc vi} line at 33.736~\AA. The third-order \ion{Ne}{\sc x} 
resonance line at 12.134~\AA\ is blended by the \ion{S}{\sc xii} transition at 36.500~\AA, 
and the second-order \ion{O}{\sc viii} 
resonance line contributes to the wing of the \ion{S}{\sc xi} and \ion{Si}{\sc xiii} feature at 37.787~\AA.
Some free unblended higher-order lines are given in Table~\ref{higherOrderLines}. The higher
order lines are implemented in the model fits of Sects.~5.1 and 5.2.
First-order lines in the wavelength range above 50~\AA\ belong to lower ionization stages 
of silicon and iron, and their optimal ionization temperature is around 0.1~keV.  

Some stronger lines are collected in Table~\ref{lineflux2} and 
shown in Fig.~\ref{spectrallines}. 
This table shows the 
line shifts and the line broadenings. The line shifts show velocities between -160 and +160~km~s$^{-1}$. The
earth's orbital velocity in the direction of \object{$\delta$~Ori} during the observation and 
the velocity of \object{$\delta$~Ori} itself (Wilson 1963) are both about 16~km~s$^{-1}$,
but they have opposite signs and are canceling.
The obtained velocities are comparable with those given by Miller et al. (2002), based on observations by {\it HETGS}. 
 The HWHMs of the strong lines result in a 
velocity of 
about 840~km~s$^{-1}$, below the terminal velocity of 2000~km~s$^{-1}$. Our value is higher than the values 
given by Miller et al. (2002), based on {\it HETGS}. This might be due to an inconsistency between the instrumental LSFs of {\it LETGS} and {\it HETGS}. However, our value agrees with Wojdowski \& Schulz (2005), 
who also used {\it HETGS}.
The ultimate velocity at the end of the wing, corrected for the instrumental width are also given. These values are about 1850~km~s$^{-1}$ with an uncertainty of 200~km~s$^{-1}$.
The lines in the separated spectra obtained from the 
two time intervals (0-49~ks and 49-98~ks) were measured, and
no significant differences between the line measurements of the two time intervals have been established.

\begin{table*}[bh!]
\caption[]{Wavelength shifts and half width at half maximum (HWHM) values for some strong isolated 
lines in the spectrum of $\delta$~Ori The values have been compared with those from Miller et al. 2002.
}
\label{lineflux2}

\begin{tabular}{l@{\ }l@{\ }c@{\ }c@{\ }c@{\ }c@{\ }c@{\ }c@{\ }c@{\ }c@{\ }}		 
		   \hline \hline

ion          & $\lambda_0^a$& $\lambda_{obs}$& $\Delta \lambda$&shift&shift(Miller) &  HWHM&HWHM(Miller)&blue wing&red wing\cr 
             & [\AA]      & [\AA]        &[\AA]            &  in km~$s^{-1}$&  in km~$s^{-1}$ &in km~$s^{-1}$&in km~$s^{-1}$&in km~$s^{-1}$&in km~$s^{-1}$\cr 
\hline

Ne\,{\sc x}	&12.134 &~~12.1305(.0030)&~~~-0.0035(.0030) &~~~~-87(~74)&~~~-150(100)&~653(123) &~~~420(170)&~~-1168&1246\cr
Fe\,{\sc xvii}  &15.015 &~~15.0100(.0023)&~~~-0.0050(.0023) &~~~-100(~46)&~~~~-50(100)&~840(~80) &~~~510(220)&~~-1734&1915\cr
Fe\,{\sc xvii}  &16.777 &~~16.7745(.0034)&~~~-0.0025(.0034) &~~~~-45(~57)&~~~~100(150)&~853(~89) &~~~420(250)&~~-1804&1783\cr
O\,{\sc viii}	&18.969 &~~18.9684(.0015)&~~~-0.0006(.0015) &~~~~~-9(~24)&~~~~~60(100)&~854(~30) &~~~700(100)&~~-2009&2067\cr
O\,{\sc vii}    &21.602 &~~21.6148(.0038)&~~~~0.0128(.0038) &~~~ 178(~53)&~~~~~-------&~792(~47) &~~~~--------&~~-1808&1826\cr
O\,{\sc vii}    &21.804 &~~21.7930(.0041)&~~~-0.0110(.0041) &~~~-151(~56)&~~~~~-------&~812(~61) &~~~~--------&~~-1915&2066\cr
N\,{\sc vii}    &24.781 &~~24.7866(.0100)&~~~~0.0056(.0100) &~~~~~68(121)&~~~~120(400)&~750(169) &~~~420(870)&~~-1704&1669\cr
C\,{\sc vi}     &33.736 &~~33.7252(.0050)&~~~-0.0108(.0050) &~~~~-96(~44)&~~~~~-------&~835(~55) &~~~~--------&~~-1834&1751\cr

\hline
\end{tabular}
\begin{flushleft}
{
Notes:\\
$a$: $\lambda_0$ are the theoretical wavelength values from Kelly (1987).\\
}
\end{flushleft}

\end{table*}

\section{He-like line ratios}
The individual line fluxes of the line triplets in the He-like ions, consisting of a resonance line~(r), 
an intercombination line~(i), and a forbidden line~(f), are tools for density and temperature
 diagnostics (Gabriel and Jordan 1969), because the
upper level of the forbidden line ($1s2s~^3S_1$) is depopulated by collisions in favor of the upper level of the intercombination
line ($1s2p~^3P_{1}$). On the other hand, the same effect of depopulation can be caused by a strong UV radiation 
field (Blumenthal et al. 1972). The latter gives the size of the UV field flux and therewith an indication of the distance to the stellar
surface where the X-ray emitting ions are formed. While for late type stars
the collisions and therewith the electron density are the dominant processes, for hot, early type stars the strong UV radiation field (the
location relative to the stellar surface) is most important. This has become
a well-proven diagnostic tool for establishing the radial location of the X-ray source in the stellar wind of hot stars
(Miller et al. 2002; Leutenegger et al. 2006; Waldron \& Cassinelli 2007).
The appropriate features in the He-like ions of \ion{C}{\sc v},
 \ion{N}{\sc vi}, \ion{O}{\sc vii}, \ion{Ne}{\sc ix},
\ion{Mg}{\sc xi}, and \ion{Si}{\sc xiii} lie in the range from 5 to 45 \AA, which is covered by {\it LETGS}. 

\begin{table}[h!]
\caption[]{He-like and H-like over He-like line flux ratios and corresponding distances in $R_*$ 
and temperatures. Values obtained by Miller et al (2002) are also given.}
\label{fi-ratios}

\begin{tabular}{l@{\ }l@{\ }l@{\ }l@{\ }}		 
		   \hline \hline

                & this work  &       &Miller et al. 2002\cr 
\hline
ion             & $f/i$      &distance($R_*$)&$f/i$ \cr 
\hline
C\,{\sc v}      & 0.31(.10)  & 75(14)  &---- \cr
N\,{\sc vi}     & 0.38(.17)  & 49(13)  &---- \cr
O\,{\sc vii}    & 0.052(.016)& 9.0(1.5)&$\la$ 0.2\cr
Ne\,{\sc ix}    & 0.10(.06)  & 3.4(1.1)&$\la$ 0.1 \cr
Mg\,{\sc xi}    & 0.73(.48)  & 3.2(1.4)& 0.5(.4)\cr
Si\,{\sc xiii}  & $\la$ 0.6  &  ---    & 2.2(1.5)\cr

\hline
ion             & $(f+i)/r$    &T(MK)&$f+i/r$ \cr 
\hline
C\,{\sc v}      & 0.96(.34)  & 1.0   &---- \cr
N\,{\sc vi}     & 0.86(.23)  & 1.6   &---- \cr
O\,{\sc vii}    & 0.93(.06)  & 1.7   &0.78(0.35)\cr
Ne\,{\sc ix}    & 0.94(.14)  & 2.3   &1.09(0.44)\cr
Mg\,{\sc xi}    & 1.08(.37)  & 2.1   &0.54(0.28)\cr
Mg\,{\sc xi}$^a$& 0.81(.23)  & 4.4   &----\cr
Si\,{\sc xiii}  & 0.69(.27)  & 8.0   &0.84(0.34)\cr
Si\,{\sc xiii}$^a$& 0.77(.22)& 6.4   &----\cr

\hline
ion             & $r_H^b/(r+i+f)$&T(MK)&\cr 
\hline
C\,{\sc vi/v}     & 0.91(.18)& 1.18(0.07)& \cr
N\,{\sc vii/vi}   & 0.55(.09)& 1.64(0.08)& \cr
O\,{\sc viii/vii} & 0.55(.02)& 2.44(0.02)&\cr
Ne\,{\sc x/ix}    & 0.35(.04)& 3.8(0.1)&\cr
Mg\,{\sc xii/xi}  & 0.24(.09)& 6.0(0.5)&\cr
Si\,{\sc xiv/xiii}& 0.22(.14)& 9.4(1.3)&\cr

\hline

\end{tabular}
\begin{flushleft}
{
Notes:\\
$a$: Average of our value and the {\it HETGS} value obtained by Miller et al. (2002)\\
$b$: $r_H$ is the line flux for the H-like resonance line doublet
}

\end{flushleft}

\end{table}

The line fluxes of these special cases have been measured a slightly different way from the other lines 
given in Table~\ref{lineflux}.
After the final overall multitemperature CIE-model fit (see Sect.~5), 
limited wavelength ranges were selected just around the relevant 
line features of the He-like 
ions. Then the lines from the ions under investigation were ignored from the model. In this way 
the continuum is based on the multitemperature fit and possible blends are taken into account. 
From that point on the same procedure was used as in
Sect.~4: a Gaussian was folded through the response matrix and fitted to the line profiles of each ion line individually.
The line flux ratios (f/i) obtained in this way are collected in Table~\ref{fi-ratios}. They might differ from ratios derived from 
Table~\ref{lineflux}, as in that table total feature fluxes, polluted by blends and satellite lines are given 
with their dominant identifications. The line flux ratios are compared with those given by Miller et al. (2002) using {\it HETGS}. Their values
and ours are in good agreement. 
The ratios have been
applied to a Planck UV-radiation field of 33000~K at the wavelengths of the 1s2s$^3$S -- 1s2p$^3$P transitions of 
the corresponding ions, in combination with scaled formulas given by Blumenthal et al (1972). This UV radiation field decreases with distance from the stellar surface by a dilution 
factor:
\begin{equation}
W(r)=\frac{1}{2}~\left[1-\left(1-\left(\frac{R_{*}}{r}\right)^2\right)^{1/2}\right].
\end{equation}
As a result of the radial dependence of the radiation field, the observed $f/i$ ratio can be used to derive
the radial location of the He-like ions that are producing the observed $fir$ lines.

\begin{figure}[h!]
\hbox
{\includegraphics[width=\columnwidth]{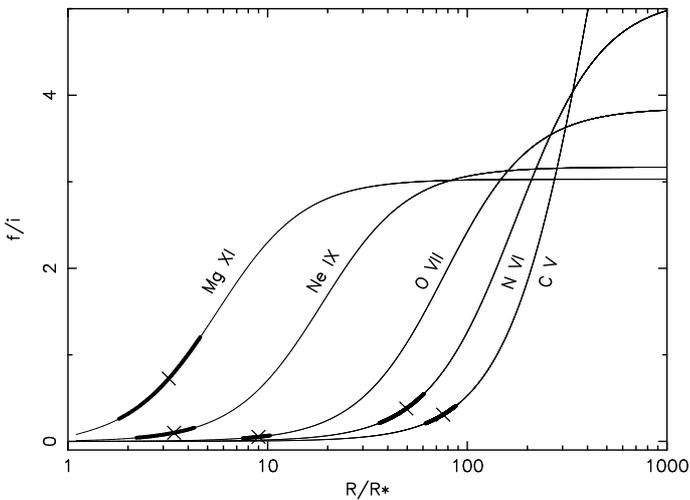}}
\caption{The $f/i$ ratio versus the distance to the stellar surface in stellar
radii (R*).}
\label{fi}
\end{figure} 

The influence of the UV radiation field on the population of the $1s2s~^3S_1$ and $1s2p~^3P_1$ levels is reflected in the f/i ratio,
hence in the distance to the stellar surface as shown in Fig.~\ref{fi}. 
The shapes of the curves are the same as those given by Miller et al. (2002) for \ion{Mg}{\sc xi}, 
\ion{Ne}{\sc ix}, and \ion{O}{\sc vii}. The curves of \ion{N}{\sc vi} and \ion{C}{\sc v} are new and established as a result of
our investigation. 
The bold intervals on the trajectory show the average distance from the stellar surface 
(including the $1\sigma$ statistical error) where the ion is formed. From this figure it is clear that the hotter 
ions (\ion{Mg}{\sc xi} and
\ion{Ne}{\sc ix}) are formed closer to the stellar surface than the cooler ones (\ion{C}{\sc v} and \ion{N}{\sc vi}). 
In the short wavelengths our
results are in good agreement with Fig.~4 in the paper by Miller et al. (2002) and with the values of \ion{Mg}{\sc xi} and
\ion{O}{\sc vii} by Waldron \& Cassinelli (2007),
 based on {\it HETGS} observations. 
At the longer wavelengths, 
\ion{N}{\sc vi} and \ion{C}{\sc v} (outside the {\it HETGS} range) show that the X-rays are formed 
much farther out in the wind than the wind
acceleration zone, which stops at about 10~R$_*$. The \ion{N}{\sc vi} is located at 49(13)$R_*$ and \ion{C}{\sc v}
 at 75(14)$R_*$ with temperatures of 1.6~MK and 1.0~MK respectively (see Table~\ref{fi-ratios}). 

Only an upper value has been determined for \ion{Si}{\sc xiii}. 
This feature is better imaged by {\it HETGS}, and
therefore the distance of its plasma to the stellar surface is  excluded from Table~\ref{fi-ratios} and Fig.~\ref{fi}.
Also included in Table~\ref{fi-ratios} are the temperature indicators $G=(f+i)/r$ of the He-like ions, as well as the
ratios of the H-like resonance lines and the He-like line triplets. They result in temperatures for the X-ray emitting 
plasma, related to the location of the specific ions in the plasma.

\section{Multitemperature fitting}

Along with the individual line analysis we investigated the LETG-spectrum of \object{$\delta$ Ori} using 
a multithermal model 
for optically thin plasma in CIE
as implemented in SPEX (Kaastra et al. 1996a) in combination with MEKAL
(Mewe et al. 1985, 1995; Kaastra et al. 1996b)\footnote{\textrm {http://www.sron.nl/SPEX}}. 
It provides a calculation with thousands of lines and a model electron continuum.
The ionization equilibrium is based on calculations by Arnaud \& Rothenflug (1985) and Arnaud \& Raymond (1992), the latter 
especially for iron. Application of a
 non equilibrium ionization model did not improve the description of the spectrum.  

 \begin{table}[ht!]
 \caption{The 3-T fit to the {\it LETGS} spectrum of \object{$\delta$ Ori}, applying a CIE-model to the total (time)
 spectrum. The distance is put at 212~pc (Van Leeuwen 2007).} 
  \label{multiT-fit2}
  \begin{center}
    \leavevmode
    \footnotesize
    \begin{tabular}[h]{lc}
    \hline
      \hline
      Parameters &  \\
      \hline
      model              &  \\ 

      logN$_H$[cm$^{-2}]$& 20.18(.03)\\ 
      z (Redshift)       & -2.3(.5)e-4 \\
      v$_{mic}$          &1039(21) \\      
      $T_1$[keV]         & 0.098(.005) \\
      $T_2$[keV]         & 0.228(.008)\\
      $T_3$[keV]         & 0.647(.010)\\
      \hline      
      $EM_1$[10$^{54}$cm$^{-3}$] &2.05(.20)\\
      $EM_2$[10$^{54}$cm$^{-3}$] &2.27(.09)\\
      $EM_3$[10$^{54}$cm$^{-3}$] &0.88(.04)\\

      \hline      
      $L_1$[10$^{31}$erg/s]$^a$     & 7.10\\
      $L_2$[10$^{31}$erg/s]         & 5.14\\
      $L_3$[10$^{31}$erg/s]         & 2.59\\
      $L_{Tot}$[10$^{31}$erg/s]     &14.83\\

      \hline    
      $Nh_1$[10$^{20}$cm$^{-2}$] &0.21(.16)\\
      $Nh_2$[10$^{20}$cm$^{-2}$] &1.52(.37)\\
      $Nh_3$[10$^{20}$cm$^{-2}$] &26.7(4.0)\\

      \hline      
      C                       & 1.080(.016)\\
      N                       & 1.046(.020)\\
      O                       & 0.859(.009)\\
      Ne                      & 1.243(.015)\\
      Mg                      & 1.650(.022)\\
      Si                      & 0.913(.012)\\
      Fe                      & 1.000(fix) \\
      $C-stat$/d.o.f.         & 2836/1609 \\
      $C-stat_{red}$          & 1.76	   \\
      
      \hline 
      \end{tabular}
  \end{center}
\begin{flushleft}
$a$: In the range from 0.07 to 3 keV. ({\it LETGS} band).\\
\end{flushleft}
  
\end{table}

The fits were applied using the C-statistics approach. For the transition probabilities of the 
model line features, an uncertainty of 15\% was assumed, a value that accords with the {\it NIST}~database 
\footnote{\textrm {http://www.nist.gov/pml/data/asd.cfm}}. The first calculation was fitted to the data, 
taking the errors on the data count rate into account. In the follow up of the fitting procedure the errors on the model
count rate were applied to avoid over-influences of incidental low count rate bins.

\subsection{A 3-T model}
 
We started the fitting procedure with more fixed temperature components, out of which only three components were significant.
From that point on a three temperature (3-T) fit was applied for the CIE-model.
The values of the fit are collected in Table~\ref{multiT-fit2}. This table shows the interstellar column density N$_H$, 
the obtained temperatures, the emission measures, the luminosities of the
different temperature bins, absorption column densities for the three temperature components in the wind, 
abundances, line shifts (z), and line broadening (vmic).

Three statistically well-determined temperature components at 0.10, 0.23, and 0.65~keV have been established.
The two coolest components correspond to the lines of the He-like ions, discussed in Sect.~4. According to
conclusions from that section, the hottest component (0.65~keV), that is responsible for \ion{Si}{\sc xiv} and \ion{S }{\sc x v} is formed deeper in the wind.

\begin{figure}[h!]
\hbox
{\includegraphics[width=\columnwidth,height=6.5truecm]{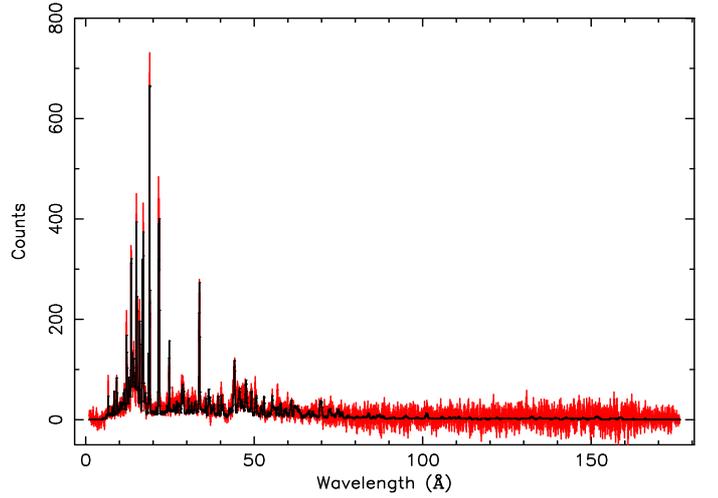}}
\caption{The spectrum of \object{$\delta$ Ori} (red/gray), together with the model (black) derived from the multitemperature fit 
applying the parameters given in Table~\ref{multiT-fit2}.}
\label{spectralfit} 
\end{figure}

\begin{figure}[h!]
\hbox
{\includegraphics[width=\columnwidth,height=5.5truecm]{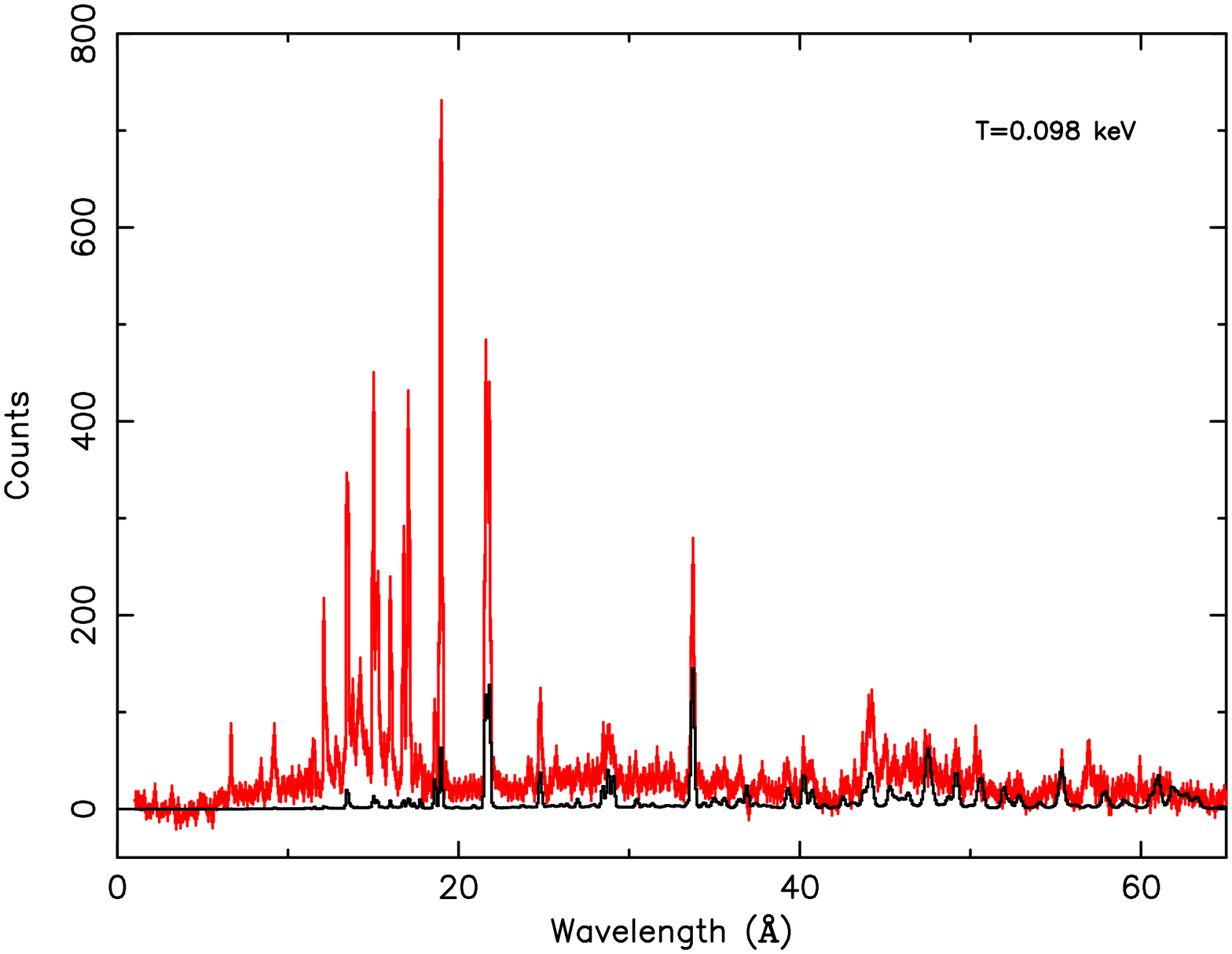}}
{\includegraphics[width=\columnwidth,height=5.5truecm]{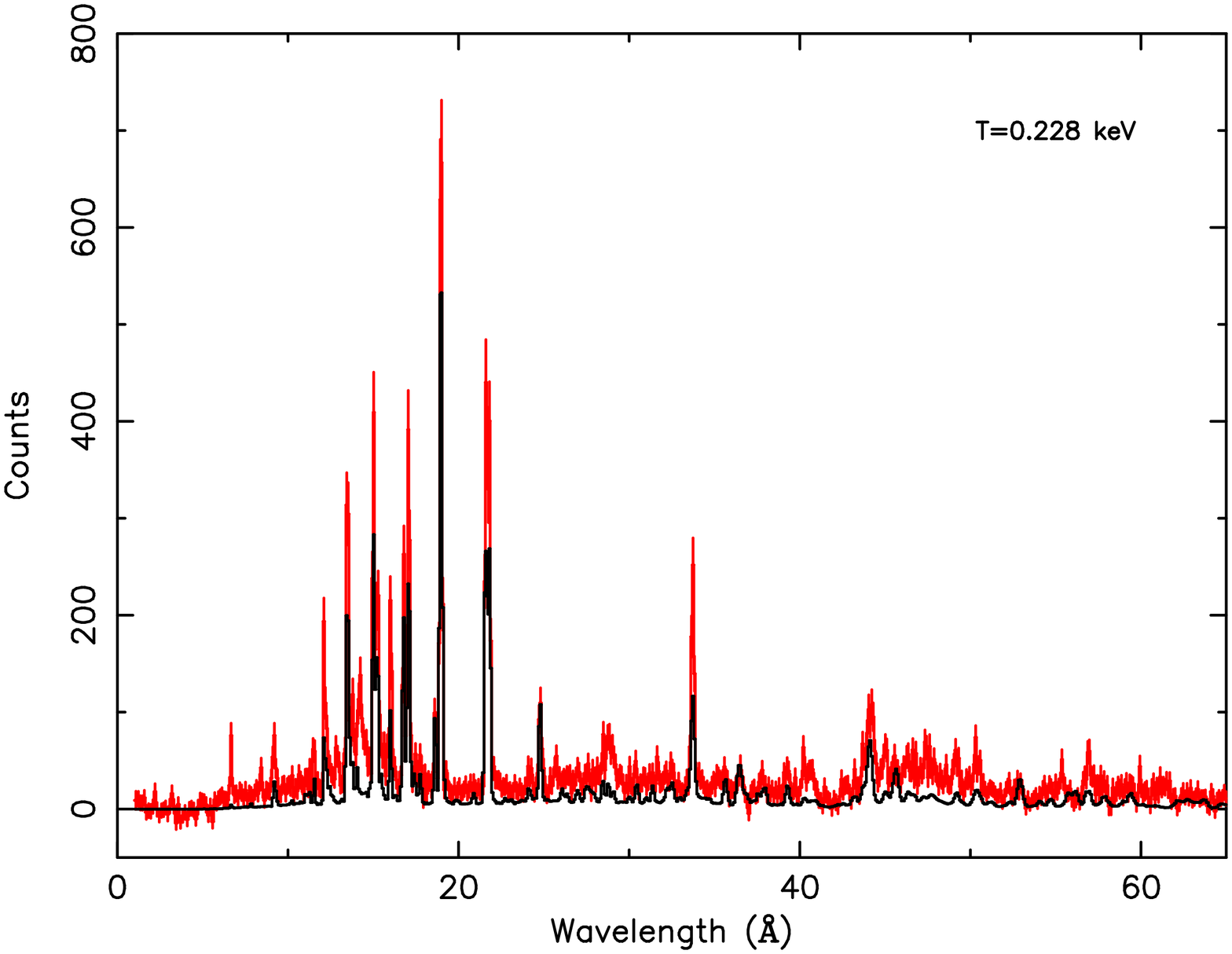}}
{\includegraphics[width=\columnwidth,height=5.5truecm]{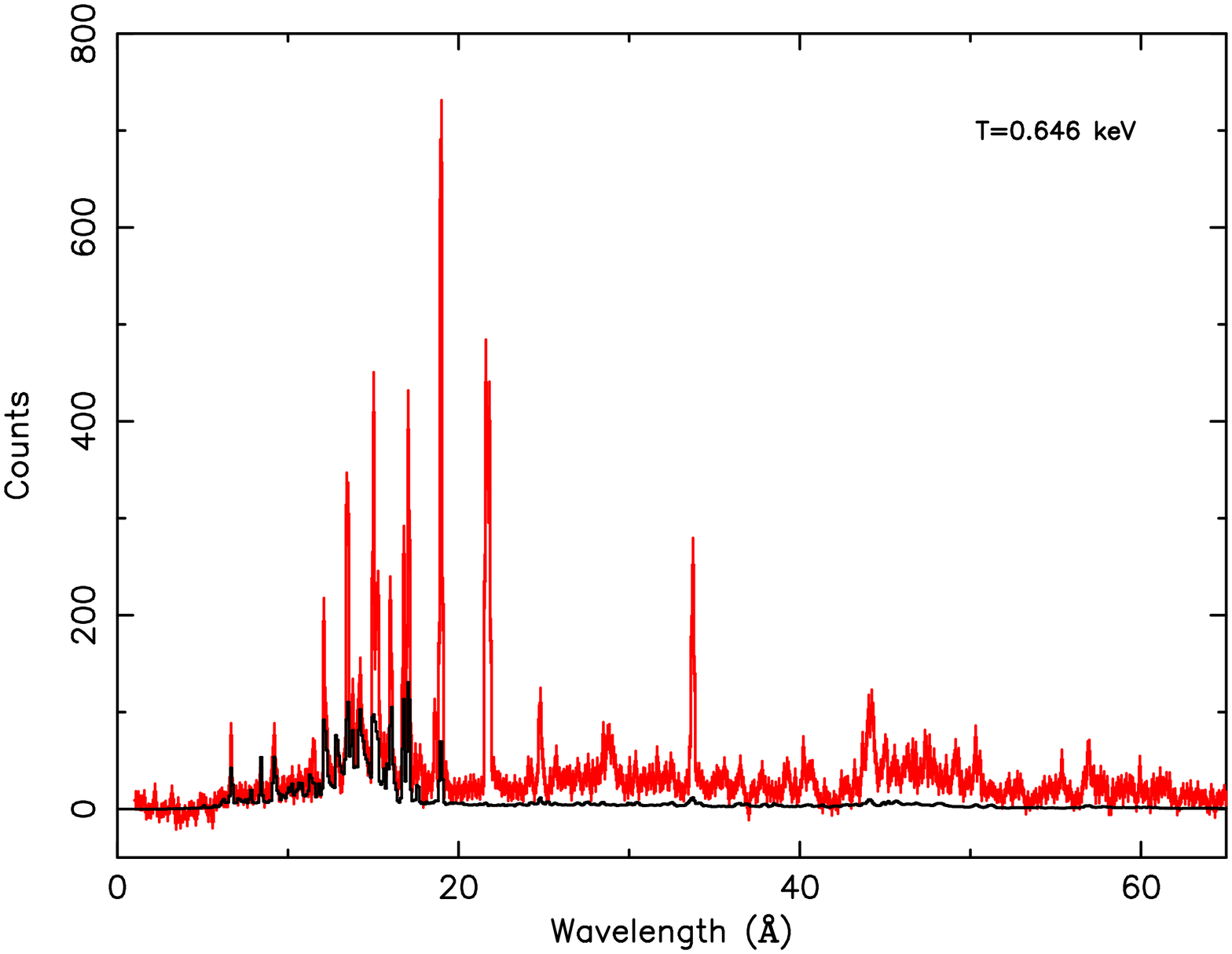}}
{\includegraphics[width=\columnwidth,height=5.5truecm]{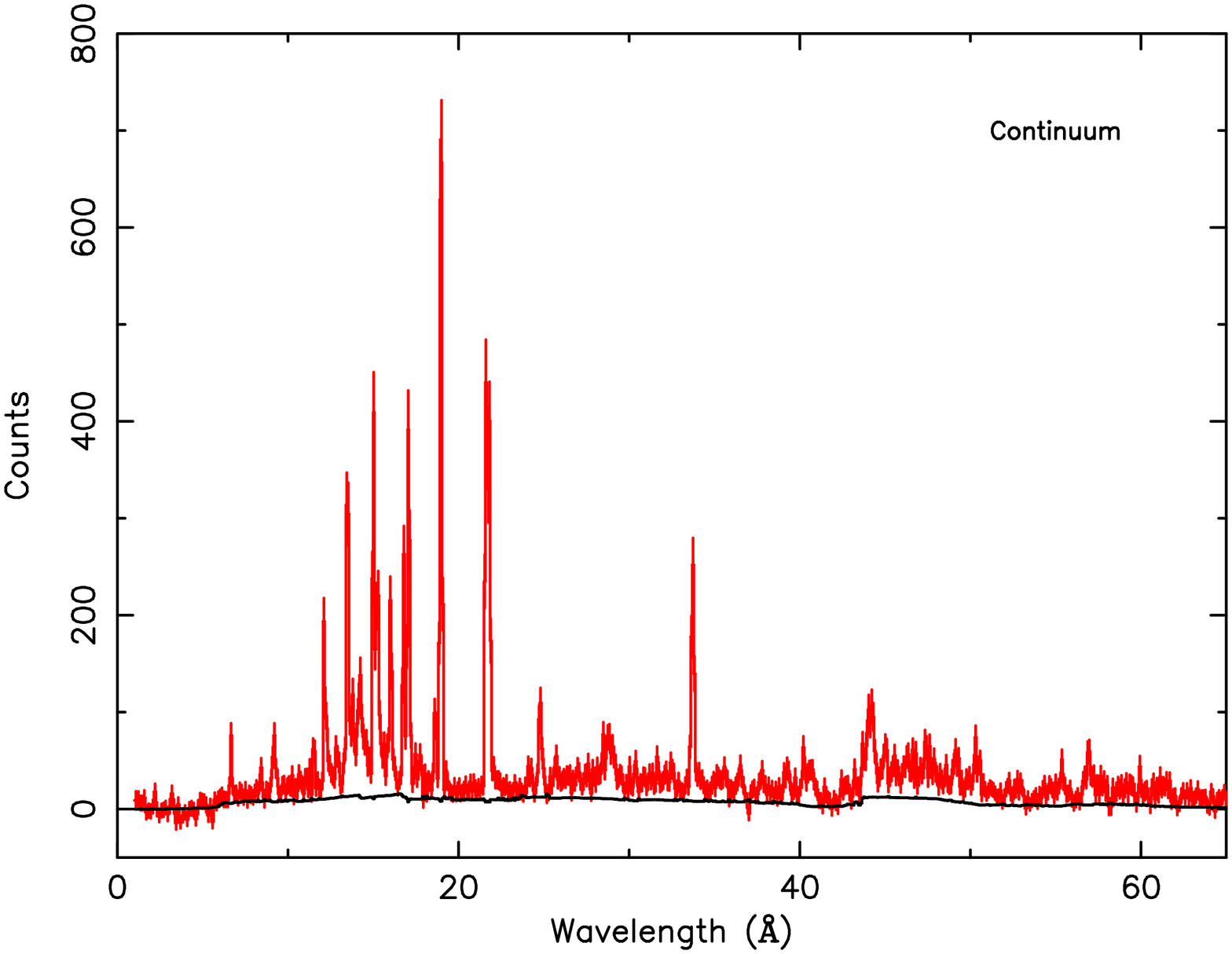}}
\caption{The spectrum of \object{$\delta$ Ori} (red/gray), together with the three individual temperature contributions 
to the model (black), applying the parameters given in Table~\ref{multiT-fit2}. They imply the line and continuum
radiation from the individual components. The bottom panel shows the
model electron continuum (black). This implies the sum of the three continua.}
\label{spectralfitCIE} 
\end{figure}

The results of the total 3-T fit to the spectrum are shown
in Fig.~\ref{spectralfit}. 
The lines and continua from the three individual temperature components
are shown in the three top panels of Fig.~\ref{spectralfitCIE}. The bottom panel shows the data with the calculated continuum as sum of the three continua of these components.

The tabulated emission measure is defined as 
\begin{equation}
EM=n_en_HV, 
\end{equation}
in which $n_H$=0.85$\times$$n_e$.
 
Knowledge of the electron density offers the possibility to calculate the X-ray
emitting volume. The electron density for hot stars with a spherically symmetric stellar wind can be obtained via the following continuity equation:

\begin{equation}
\dot M = 4\pi R^2 \rho(R) V(R).
\end{equation}

In this formula $\dot M$ is the mass loss rate (1.07$ \times 10^{-6} M_\odot$ /year (Lamers \& Leitherer 1993)) for
\object{$\delta$ Ori}, 
R is the distance from the stellar surface, $\rho$(R) the density at distance R, and V(R) the wind velocity at distance R, related to the
terminal velocity ($V_{\infty}$) by
\begin{equation}
V(R) = V_{\infty}\times (1 - R_*/R)^{\beta},
\end{equation}
in which $\beta$ is a coefficient between 0.85 for subgiants and 1.1 for supergiants. We use a value of 1.0 for $\beta$.  

The InterStellar Medium (ISM) N$_H$ was fitted and found to be 1.5$\times 10^{20} cm^{-2}$.
Three additional wind absorption parameters were applied using the hot module in {\it SPEX}. It describes a warm
plasma, as the CIE-model, while calculating the continuum absorption, as well as the line absorption. Emission lines
with the highest oscillator strength are mostly influenced. The temperatures of this hot plasma were coupled
to the temperatures of the components in the CIE-model. The hottest component, which is formed deeper in the wind (see Sect.~4), results in the highest absorption column density. No absorption for the cool wind (33000~K) could be determined. 
This implies that the overall cool stellar wind absorption is not taken into account and might have been 
taken over partly by the fitted
ISM-parameter N$_H$.

A slight shift in $\lambda$ ($\approx$~-2.3e-4) was applied to optimize the fitted wavelengths. A line broadening parameter v$_{mic}$
of 1039 km/s was added and fitted to the overall spectrum to take into account the broadening of the line features.
The HWHM of a line is related to the v$_{mic}$ by the following formula (see also Sect.~3):

\begin{equation}
HWHM/\lambda_{obs} =\frac{\sqrt{\ln 2}v_{mic}}{c}.
\end{equation}

The X-ray luminosities of the three individual temperature components are given over the LETG energy range from 0.07 to 3.0~keV. 
For the luminosities no individual
statistical errors have been determined. Their uncertainties are supposed to have the same relative values as those
for the emission measures because the luminosity is mostly determined by the emission measures in this temperature
range, and therefore $L_ X \propto EM$. 
The total luminosities (summed over the three temperature components) in other energy bands 
(0.5-4~keV, 0.1-2.4~keV, and 0.4-4~keV) 
for {\it Einstein, ROSAT, ASCA} 
given in the papers by Cassinelli \& Swank (1983), Haberl \& White (1993), and 
Corcoran et al. (1994), respectively, do agree with our values when taking differences in interstellar absorption or distance into account.

The abundances are relative to solar photospheric values (Lodders et al. 2009).
The absolute values of the abundances are strongly influenced by the obtained values 
of the emission measures. The product of abundances and emission measures (A $\times$ EM), however, is the robust quantity
for the line spectrum, therefore, abundance-ratios are often given relative to an abundant element in the stellar atmosphere.
Here the abundance of iron was fixed at its solar photospheric value.
The abundance values agree with the value of
$\approx$~0.9 that Sim\'on-D\'iaz (2010) finds for oxygen and silicon abundances in the
Orion star-forming region. 

A fit to the spectrum taken by {\it HETGS} applying the same set of parameters needs a lower value 
for v$_{mic}$ (see also Sect.~3). This might be due to inconsistency between the LSFs of the used instruments. 
Also the description of the \ion{Si}{\sc xiii} and \ion{Si}{\sc xiv} line features is not satisfactory. 
This might be because these features are near the end of the calibrated area for both instruments.

To investigate the time dependence of the spectrum, as suggested by the light curve (see Fig~\ref{lightcurve}) we split
 the observation into two time intervals, 0-49~ks and 49-98~ks.
The emission measure of the hottest component (formed closest to the stellar surface) is about 10\% lower in the first
time interval than in the last 49~ks. This might agree with some eclipsing by the companion star. 

\subsection{DEM modeling}
\begin{figure}[h!!]
\hbox
{\includegraphics[width=\columnwidth]{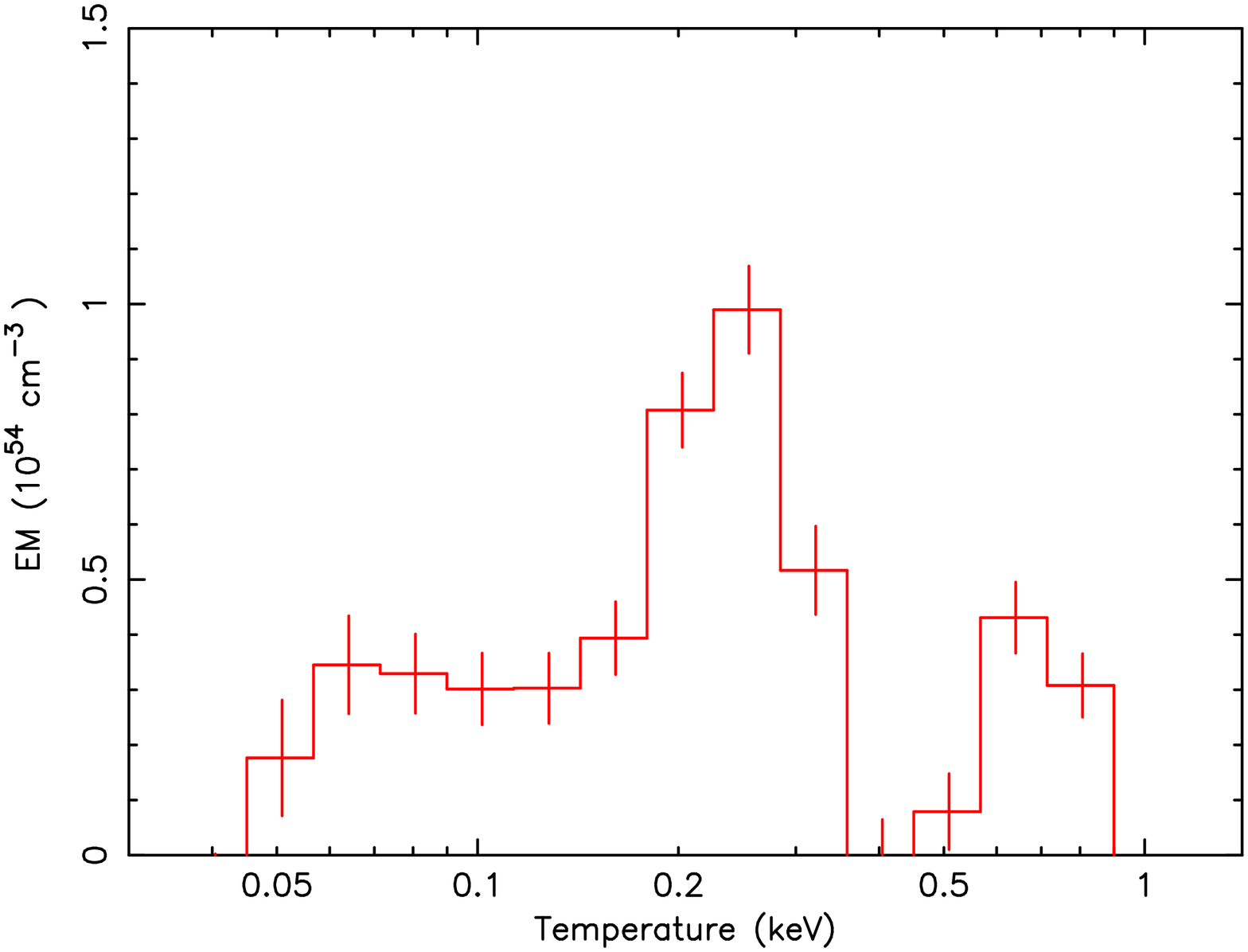}}
\caption{DEM-modeling of \object{$\delta$~Ori}, based on the regularization method (Kaastra et al. 1996b) (see text). 
The $EM$ is given in units
of $10^{54}$cm$^{-3}$.
}
\label{demreg}
\end{figure}

To show the smooth connection between the separated temperature bins, we
performed a  continuous DEM-modeling for the
spectrum of \object{$\delta$~Ori} by applying the regularization method (Kaastra et al. 1996b) in the
SPEX-code. This method uses direct matrix inversion with the additional constraint that the
second-order derivative of the solution with respect to the temperature is as smooth as possible.
In this DEM~modeling the abundances given in Table~\ref{multiT-fit2} have been used. The results are shown in
Fig.~\ref{demreg}.  

\begin{figure}[h!!]
\hbox
{\includegraphics[width=\columnwidth]{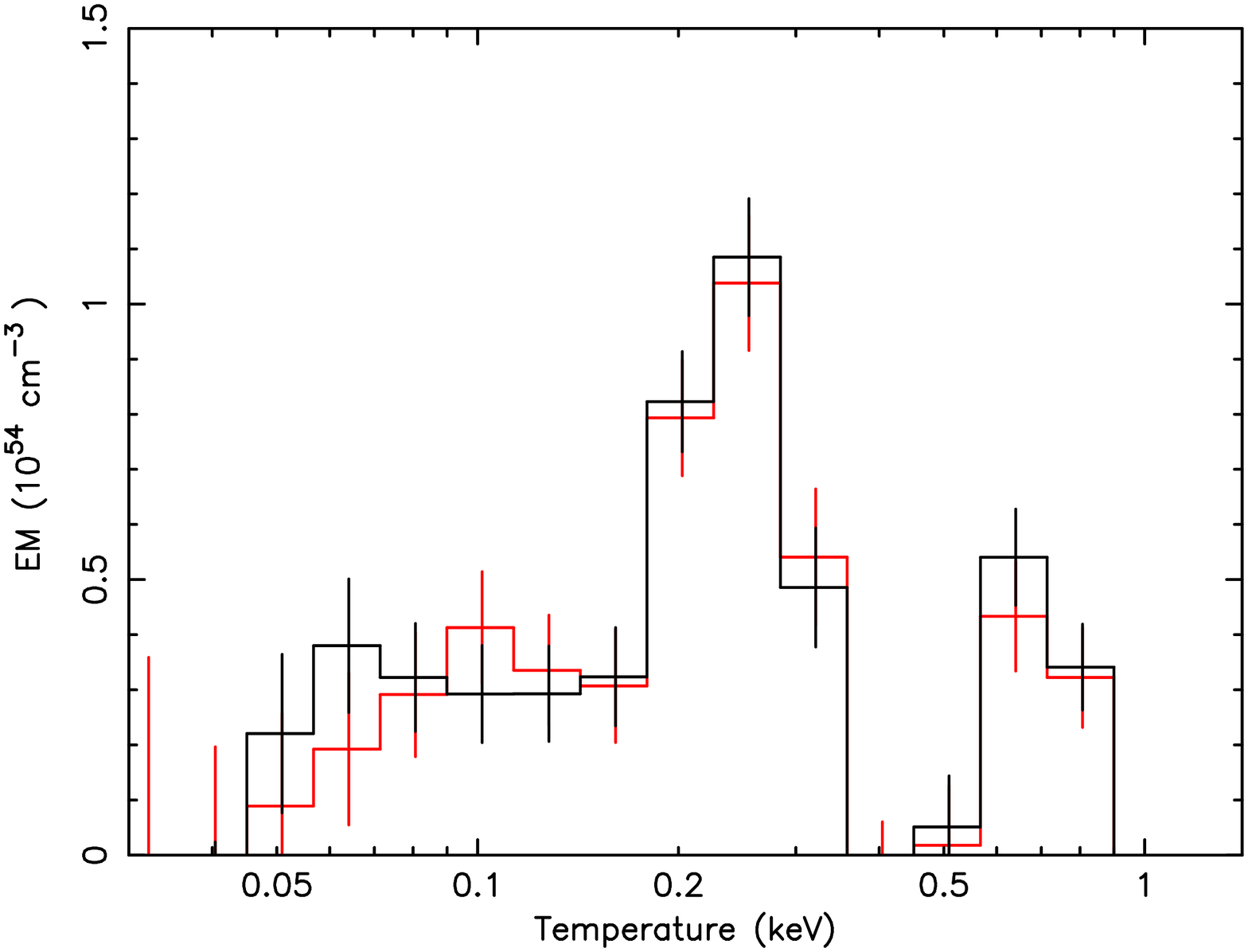}}
\caption{DEM-modeling of \object{$\delta$~Ori}, based on the regularization method. The two DEMs show a modeling of the 
spectrum over the time interval 0-49~ks (red/gray) and  49-98~ks (black).
}
\label{demreg2}
\end{figure}

The results of the fits, collected in Table~\ref{multiT-fit2} and Fig.~\ref{demreg}, are in good agreement.
The peak of our DEM agrees with the peaks found by Miller et al. (2002) and Wojdowski \& Schulz (2005).
However, our DEM extends more to lower temperatures and their values extend further to higher temperatures as a consequence of using two different instruments. The peak value given by Wojdowski \& Schulz (2005) is higher.

As for the CIE-fit we also performed a DEM-modeling for the partial
spectra of 0-49~ks and 49-98~ks, again applying the regularization method. These two DEM-modelings are shown
in Fig.~\ref{demreg2}. Although the hottest component is again a bit stronger for the last 49~ks, no significant difference can be established between fits to the partial spectra.

\section{Discussion and conclusions}
The X-ray spectrum of the O-star \object{$\delta$ Ori} obtained with {\it LETGS} is rich in lines belonging to H-like and He-like
ions of C, N, O, Ne, Mg, and Si. A weak He-like (\ion{S}{\sc xv}) triplet
is also present.
Other possible weak features present below 5~\AA\ fall outside the calibrated bandwidth of the {\it LETGS}.  
Iron L-shell lines are dominant. The Ne-like transitions, between electron configurations of an inert noble gas,
 are favored for iron (Fe XVII) as
well as for calcium (Ca XI), and nickel (Ni XIX).

Above 50~\AA\ a few new line features were identified, some of which are third-order lines associated with strong 
first-order features
located between 15 and 35~\AA. No lines of cool ions were detected at wavelengths higher than 70~\AA. 
The main reasons for the absence of these line features is the absence of a noticeable volume of cool (0.05~keV) plasma
and the interstellar absorption.

Based on the forbidden line over intercombination line ratio in He-like ions the average distance of the X-ray emitting
plasma to the stellar surface has been determined. Depending on the ion (temperature), the plasma emitting region is between 2
and 80~$R_*$ for ions from \ion{Mg}{\sc xi} to \ion{C}{\sc v}. The cooler plasma (lower ionization stages) is located farther
out in the wind. For the hot ions, \ion{Mg}{\sc xi}, \ion{Ne}{\sc ix}, and  \ion{O}{\sc vii}, the average X-ray emitting
plasma is from 2 to 10~$R_*$, respectively. This is in accordance with observations by Waldron \& Cassinelli (2007) 
and Miller et al. (2002) 
using {\it HETGS}. Using {\it LETGS} that extends to longer wavelengths, the cooler ions 
\ion{N}{\sc vi} and  \ion{C}{\sc v} have been observed as part of this investigation. The X-rays from these ions 
are formed on average at 50~$R_*$ and 75~$R_*$, respectively,
much farther out in the wind than the acceleration zone, where the final velocity has been reached. 
Here the wind is expected to be steady. However, small
deviations in velocity (weak shocks) will be able to produce the cooler ions.

The spectrum is described well by a CIE model. 
Three well-distinguished temperature components have been 
determined: 0.10, 0.23, and 0.65~keV. The first two are in good agreement with the two temperatures 
found by Haberl \& White (1993). They did not determine a third temperature component. The two lower temperature components (0.098 and 0.228~keV)
describe the line-forming regions of the cooler ions, \ion{N}{\sc vi} and  \ion{C}{\sc v}, and the moderately ionized
species \ion{Mg}{\sc xi}, \ion{Ne}{\sc ix}, \ion{O}{\sc vii}, and \ion{O}{\sc viii}, respectively, while the hottest component is supposed to
produce even hotter ions 
deeper in the wind and a hot continuum.

Three emission measures are established, related to the obtained temperatures.  
Knowledge of the electron density offers the possibility to calculate the X-ray
emitting volume. 
Applying the distances (see Table~\ref{fi-ratios} and Fig.~\ref{fi}) and the wind velocity and density formulas 4 and 5,
given in Sect.~5.1, we have calculated the
electron densities for the regions where the specific ions produce X-ray radiation. 

Because the distances given in Table~\ref{fi-ratios} and Fig.~\ref{fi} are average values with the statistical
error on this average and not the limits of the X-ray emitting distance, it is difficult to calculate the total wind
volume in which the conditions are optimal for the corresponding He-like ion emission. 
By assuming a wind volume from half to twice the obtained average distance we get the percentage of the wind that
contributes to that ion.
The emitting volumes are about 1$\%$ of the wind volume for the cooler plasma and less for the hotter ions.

The N$_H$ of the ISM is mainly based on the long wavelength part of the 
spectrum. Its value of 1.51(.06)$\times 10^{20} cm^{-2}$ is in good agreement with 
the value of 1.56$\times 10^{20} cm^{-2}$ found by Jenkins et al. (1999) based on 57 IUE observations, and slightly
lower than the value found by Haberl \& White (1993) for their T=0.1~keV component.
Additional wind absorption column densities have been determined (see Table~\ref{multiT-fit2}). They are related to
the temperature components of 0.10, 0.23, and 0.65~keV, with distances to the stellar surface of about 60$R_*$, 6$R_*$,
and 2$R_*$, respectively, where the plasma is formed on average. 
In combination with formula~4 we calculated that about 7$\%$ of the wind contributes to the hot wind absorption. 
No absorption of the overall cool stellar wind could be determined. This means that this absorption component is not explicitly 
taken into account during the fitting procedure. However, the effect might have been taken over 
by the other absorption parameters.

The X-ray luminosities of the three temperature components in the total {\it LETGS} band (0.07-3~keV) are 7.10, 5.14,
and 2.59~10$^{31}$erg/s. 
Two partial spectra were derived: one from the first 49~ks and another from the last 49~ks of the observation. 
This division was suggested 
by the light curve (see Fig~\ref{lightcurve}). No significant differences between the spectra, their temperature structures, their
line shapes, or line fluxes have been established.

The obtained abundances are relative to solar photospheric values (Lodders et al. 2009). They were fit by fixing the
iron abundance at its solar photospheric value and leaving the carbon, nitrogen, oxygen, neon, magnesium, and silicon
abundances free to vary. Fixing the iron abundance was done because the abundances and emission measures are strongly
anticorrelated, especially in spectra with a weak continuum to establish the emission measure. The obtained values for oxygen and silicon are in good
agreement with values found by Sim\'on-D\'iaz (2010) in the Orion star-forming region.

The lines are symmetric and their HWHMs are about 840~km~s$^{-1}$, far below the terminal velocity of 2000~km~s$^{-1}$, but
higher than the values found by Miller et al. (2002) using {\it HETGS}. This might be due to inconsistencies between the instrumental 
line spread functions in the calibration files of {\it HETGS} and {\it LETGS}.
The ultimate velocities, measured at the end of the wings, are about 1850~km~s$^{-1}$
with an uncertainty of 200~km~s$^{-1}$.

The measured line shifts given in Table~\ref{lineflux2}, as well as the z-value given in Table~\ref{multiT-fit2},
indicate that the velocity of the radiative plasma in the direction from or to the observer is low. This is well
known for OB stars, especially among giants and supergiants (Waldron \& Cassinelli 2007). 
 
The conclusion
should be that the radiation we observe is produced in plasma that has
a very low velocity or moves perpendicular to the line of sight. 
There might be several explanations. The clumps, 
which collide with the wind and produce the X-ray
radiation, have low velocities. In that case the relative velocity of the clumps to the wind is highest far out in the
wind at terminal wind velocity. As a consequence the highest stages of ionization are produced far out in the wind, which 
contradicts our observation based on the forbidden-to-intercombination line ratios in He-like ions (see Fig.~\ref{fi}).
Additionally, after the acceleration phase one does not expect velocity differences that are too high.
The secondary star and its wind play important roles in the explanation of the \object{$\delta$ Ori} spectrum.
However, this is discussed by Miller et al. (2002) and rejected.
A third option is that the clumps screen their own X-ray radiation when they are in between 
the star and the observer. Only when the wind-clump collision is almost perpendicular to the line of sight is the
radiation observable and spread over a narrow velocity range as derived from the HWHM. The effects of 
clumping and porosity on X-Ray emission-line profiles from hot-star winds have been described extensively by Owocki \&
Cohen (2006) and Oskinova et al. (2006).

\begin{acknowledgements}

The SRON National Institute for Space Research is supported financially by NWO. 

\end{acknowledgements}

\end{document}